\documentclass[english,aps, pra, twocolumn,  superscriptaddress]{revtex4-2}
\usepackage[T1]{fontenc}
\usepackage[utf8]{inputenc}
\usepackage{color}
\usepackage{babel}
\usepackage{amsmath}
\usepackage{amssymb}
\usepackage{graphicx}
\usepackage[unicode=true,pdfusetitle,
 bookmarks=true,bookmarksnumbered=false,bookmarksopen=false,
 breaklinks=true,pdfborder={0 0 0},pdfborderstyle={},backref=false,colorlinks=true]
 {hyperref}
\hypersetup{
 colorlinks,linkcolor=blue,citecolor=blue,urlcolor=blue}
\begin{document}
\global\long\def\vect#1{\overrightarrow{\mathbf{#1}}}%

\global\long\def\abs#1{\left|#1\right|}%

\global\long\def\av#1{\left\langle #1\right\rangle }%

\global\long\def\ket#1{\left|#1\right\rangle }%

\global\long\def\bra#1{\left\langle #1\right|}%

\global\long\def\tensorproduct{\otimes}%

\global\long\def\braket#1#2{\left\langle #1\mid#2\right\rangle }%

\global\long\def\omv{\overrightarrow{\Omega}}%

\global\long\def\inf{\infty}%

\title{From Bloch Oscillations to a Steady-State Current in Strongly Biased
Mesoscopic Devices}
\author{J. M. Alendouro Pinho}
\email{up201703751@fc.up.pt}

\address{Departamento de Física e Astronomia, Faculdade de Ciências da Universidade
do Porto, Rua do Campo Alegre, s/n, 4169-007 Porto, Portugal}
\address{Centro de Física das Universidades do Minho e do Porto (CF-UM-UP)
and Laboratory of Physics for Materials and Emergent Technologies
LaPMET, University of Porto, 4169-007 Porto, Portugal}
\author{J. P. Santos Pires}
\address{Departamento de Física e Astronomia, Faculdade de Ciências da Universidade
do Porto, Rua do Campo Alegre, s/n, 4169-007 Porto, Portugal}
\address{Centro de Física das Universidades do Minho e do Porto (CF-UM-UP)
and Laboratory of Physics for Materials and Emergent Technologies
LaPMET, University of Porto, 4169-007 Porto, Portugal}
\author{S. M. João}
\address{Departamento de Física e Astronomia, Faculdade de Ciências da Universidade
do Porto, Rua do Campo Alegre, s/n, 4169-007 Porto, Portugal}
\address{Centro de Física das Universidades do Minho e do Porto (CF-UM-UP)
and Laboratory of Physics for Materials and Emergent Technologies
LaPMET, University of Porto, 4169-007 Porto, Portugal}
\address{Department of Materials, Imperial College London, South Kensington
Campus, London SW7 2AZ, United Kingdom.}
\author{B. Amorim}
\address{Centro de Física das Universidades do Minho e do Porto (CF-UM-UP)
and Laboratory of Physics for Materials and Emergent Technologies
LaPMET, Universidade do Minho, 4710-057 Braga, Portugal}
\author{J. M. Viana Parente Lopes}
\email{jlopes@fc.up.pt}

\address{Departamento de Física e Astronomia, Faculdade de Ciências da Universidade
do Porto, Rua do Campo Alegre, s/n, 4169-007 Porto, Portugal}
\address{Centro de Física das Universidades do Minho e do Porto (CF-UM-UP)
and Laboratory of Physics for Materials and Emergent Technologies
LaPMET, University of Porto, 4169-007 Porto, Portugal}
\begin{abstract}
It has long been known that quantum particles moving in a periodic
lattice and subject to a constant force field undergo an oscillatory
motion that is referred to as Bloch Oscillations (BOs). However, it
is also known that, under quite general conditions, a biased mesoscopic
system connected to leads should settle in a steady-state regime characterized
by a constant electric current (described by the Landauer formula).
Since both effects are driven by a constant field, these two quantum
transport phenomena appear to be at odds with each other. Here, we
solve this apparent contradiction by theoretically demonstrating that
BOs can actually be observed in biased two-terminal mesoscopic devices
as a transient phenomenon, which relaxes for long times to a steady-state
current that agrees with the Landauer formula. Furthermore, we also
combine analytical and numerical time-evolution results for a one-dimensional
tight-binding model of a biased two-terminal mesoscopic system, in
order to characterize the decay times of the transient BOs and establish
the conditions under which they can occur.
\end{abstract}
\maketitle

\section{Introduction}

\bibpunct{[}{]}{;}{s}{}{}When electrons moving in a periodic lattice
are accelerated by a constant electric field, they give rise to oscillatory
currents. This long-established phenomenon is referred to as \textit{Bloch
oscillation}s\,\citep{Bloch29,Zener1934} (BOs) and is expected for
any quantum particle that moves across a periodic background potential
in the presence of an uniform driving force (see Glück \textit{et
al}.\,\citep{Gluck2002} for an extensive review).\,\,Despite being
theoretically well understood, the experimental observation of BOs
remains an outstanding challenge in solid-state systems\,\citep{Mendez1993}.
The fragility of electronic BOs in solid-state systems results from
the fact that their period (inversely proportional to the applied
field) is typically much larger that the electronic scattering times,
thus leading to a loss of phase-coherence before a single current
oscillation can be finalized.\,\,As such, to this day electronic
BOs have only ever been detected in synthetic semiconducting superlattices\,\citep{Bleuse1988,Feldmann1992,Waschke1993,Roskos1995}.Analogues
of BOs have been observed in a variety of alternative platforms, such
as modulated photonic waveguides\,\citep{Lenz1999,Morandotti1999,Sapienza2003,Trompeter2006,Dreisow2009,Corrielli2013},
arrays of coupled acoustic cavities\,\citep{SanchisAlepuz2007,LanzillottiKimura2010},
ultra-cold atoms in optical potentials\,\citep{BenDahan1996,Wilkinson1996,Geiger2018},
and even in superconducting \textit{q-bit} arrays\,\citep{Guo2021}.

On the other hand, it is also expected that in a mesoscopic system
connected to electrodes at different electrochemical potentials, an
electric current will begin to flow, which eventually reaches a steady-state
regime. As first argued by Landauer\,\citep{Landauer57,Landauer70},
and later generalized by Büttiker\,\citep{Buttiker86}, the steady-state
current flowing between the electrodes is proportional to the quantum
transmittance of the sample: a \textit{non-local }property that is
sample-specific and strongly depends on the precise geometry of the
device\,\citep{Fisher81,Stone88}. This result is the celebrated
\textit{Landauer formula}, which was later demonstrated\,\citep{Meir92,Wimmer2009}
to yield the same steady-state current as the one derived by Caroli
\textit{et al}.\,\citep{Caroli71}, using a non-equilibrium Green's
function formalism. It is important to note that both these approaches
assume that the system reaches a non-equilibrium steady-state, making
no attempts to describe how (or whether) this state is reached.\,\,It
was latter theoretically demonstrated that a non-equilibrium steady-state
is reached provided the electrodes have a smooth non-zero density
of states \citep{Stefanucci2004a} and that there are no bound states
in the mesoscopic device \citep{Stefanucci2007,Khosravi2009}. A smooth
density of states in the leads gives origin to a loss of memory of
the initial state of the system. Bound states, in turn, give origin
to oscillating behavior in the current. The establishment of a steady-state
in biased mesoscopic system, after an initial transient regime, has
been theoretically demonstrated in systems with and without inelastic
mechanisms, assuming that the current is driven by either the lead-sample
couplings (\textit{partitioned setup})\,\citep{Jauho94} or a static
electric field that is suddenly applied across the device (\textit{partition-free
setup})\,\citep{Cini80,Stefanucci2004a,Stefanucci2004b}.

The dynamics of current in the transient regime that precedes the
steady-state have been subject of increasing interest\,\citep{Khosravi2009,Cornean2010,Tuovinen2013,Latini2014,Tuovinen2014,Eich2016,Pal2018,Tuovinen2019,Ridley2019,Taranko2019,Pires20,Ridley2021,Ridley2022,Cao2022}.
The transient regime has been shown to unveil exotic quantum effects
that are otherwise washed out in the steady-state. Two remarkable
examples of this are:\textit{\,(i)} the ability to distinguish the
signatures of Andreev and quasi-Majorana states in quantum transport
data of superconducting nano-wires\,\citep{Tuovinen2019}, and\textit{
(ii)} the description of time-dependent radiation from biased nano-antennas\,\citep{Ridley2021}.\,\,These
studies were made possible by the recent development of numerical
time-dependent Landauer-Büttiker methods\,\citep{Bushong2005,Tuovinen2013,Popescu2016,Pal2018,Pires20,Kloss2021}.

Since the current dynamics of a biased mesoscopic system naturally
relax towards a steady-state current, the question of whether BOs
can be seen in these devices naturally arises.\,\,In a prior work,
Popescu and Croy\,\citep{Popescu2017} have theoretically shown that
persistent Bloch oscillations \textit{can occur} in mesoscopic devices
at very strong electric fields.\,\,However, in this proposal, BOs
only exist when the applied electric field is such that there is a
total reflection of electrons with no net current flowing through
the device. Hence, one might ask if this is always the case or whether
a mesoscopic device can also exhibit Bloch oscillations as a \textit{transient
regime} which eventually relaxes to a steady-state, described by the
Landauer formula.\,\,The goal of this work is to determine the conditions
in which such \textit{transient Bloch oscillations} (tBOs) are possible
in a mesoscopic device. In order to do so, we study a one dimensional
tight-binding chain, combining numerical quantum time-evolution\,\citep{Pires20,Pal2018,Suresh2022}
with quantum transmittance calculations\,\citep{Groth2014}. The
results are then physically interpreted on the basis of \textit{(i)}
Wannier-Stark localization induced by strong electric fields within
the mesoscopic sample, and \textit{(ii)} scattering states and wavefunction
matching.

The remaining of this paper is structured as follows:\,\,In Sec.\,\ref{sec:Transport-Setup-and},
we outline the model Hamiltonian considered and the numerical method
used for quantum time-evolution.\,\,The main numerical results showing
the tBO regime are presented in Sec.\,\ref{sec:From-Infinite-to}.\,\,The
decay times of the tBOs are computed within a quasiparticle approximation
in Sec.\,\ref{sec:Relating-the-Lorentzian's} .\,\,Finally,\,\,in\,\,Sec.\,\ref{sec:Conclusion}\,\,we\,\,summarize
our key findings.

\section{Model and Methods\label{sec:Transport-Setup-and}}

\subsection{Hamiltonian and initial state of the Mesoscopic Device}

\begin{figure}
\includegraphics[width=1\columnwidth]{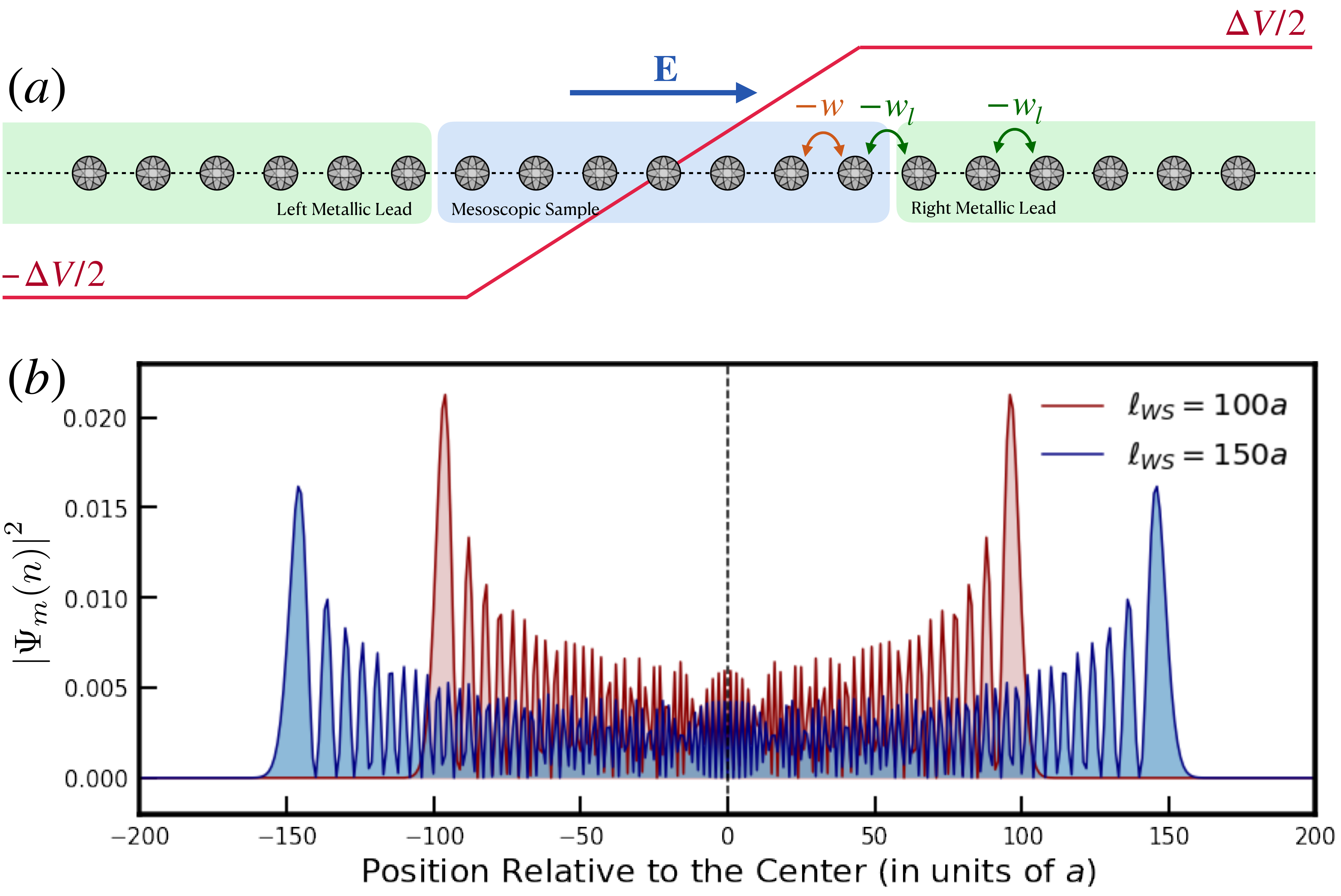}

\caption{\label{fig:models}(a) Depiction of the 1D mesoscopic device used
throughout this work. The red line represents the spacial profile
of the applied electric potential, with $\Delta V$ being the bias
voltage and $w$ ($w_{l}$) the hopping parameter inside the central
sample (each of the leads). (b) The square-modulus of a Wannier-Stark
state {[}Eq. (\ref{eq:WSS}) centered in site $m=0${]} for two values
of the applied electric field (corresponding to $\ell_{{\scriptscriptstyle \text{WS}}}=100a$
and $150a$).}
\end{figure}

We will consider transport through a one dimensional mesoscopic system,
which we described by a tight-binding model given by

\begin{equation}
\mathcal{H}(t)=\mathcal{H}_{{\scriptscriptstyle \textrm{C}}}(t)+\sum_{\alpha=L,R}\mathcal{H}_{\alpha}(t),
\end{equation}
where $\mathcal{H}_{{\scriptscriptstyle \textrm{C}}}$ describes the
central sample, $\mathcal{H}_{{\scriptscriptstyle L(R)}}$ is the
Hamiltonian for the left (right) lead, which includes the coupling
to the central region. Assuming that the central sample has $2L+1$,
the Hamiltonian of the central region reads
\begin{multline}
\mathcal{H}_{{\scriptscriptstyle \textrm{C}}}(t)=\sum_{n=-L}^{L}V_{n}^{{\scriptscriptstyle \textrm{C}}}(t)\ket n\bra n\\
-w\sum_{n=-L}^{L-1}\left(\ket n\bra{n+1}+\text{h.c.}\right),\label{eq:Hamiltonian_Central_region}
\end{multline}
where $\ket n$ describes an electron at position $n$, $w$ is the
nearest-neighbour hopping and $V_{n}^{{\scriptscriptstyle \textrm{C}}}(t)=\Theta(t)eEan$
(with $a$ the lattice spacing, $-e$ the electron charge) is the
potential due a constant electric field, applied to the central region,
that is switched on at $t=0$. The Hamiltonians of the leads read
\begin{align}
\mathcal{H}_{{\scriptscriptstyle L}}(t) & =\sum_{n=-\infty}^{-L-1}\left(V_{n}^{{\scriptscriptstyle \textrm{L}}}(t)\ket n\bra n-w_{l}\ket{n+1}\bra n+\text{h.c.}\right)\\
\mathcal{H}_{{\scriptscriptstyle R}}(t) & =\sum_{n=L+1}^{+\infty}\left(V_{n}^{{\scriptscriptstyle \textrm{R}}}(t)\ket n\bra n-w_{l}\ket{n-1}\bra n+\text{h.c.}\right)
\end{align}
where $w_{l}$ are the lead hoppings, which unlike in Refs.\,\bibpunct{[}{]}{,}{n}{}{}\citep{Popescu2017,Pal2018,Pires2019}\bibpunct{[}{]}{;}{s}{}{}
we will allow to be $w_{l}\neq w$, and $V_{n}^{{\scriptscriptstyle \textrm{L}}}(t)=-\Theta(t)\Delta V/2$,
$V_{n}^{{\scriptscriptstyle \textrm{R}}}(t)=\Theta(t)\Delta V/2$
are shits in the local energy of the lead sites, such that no electric
field is applied in the leads, with the potential difference related
to the electric field in the central region via $\Delta V=E\left(2La\right)$.
The Hamiltonian is illustrated in Fig.~\ref{fig:models}(a). In numerical
simulations, we will actually consider large, but finite leads, instead
of semi-infinite ones.

At times $t>0$ and in the limit $L\rightarrow\infty$, the Hamiltonian
$\mathcal{H}_{{\scriptscriptstyle \textrm{C}}}(t)$ reduces to the
Wannier-Stark Hamiltonian. It was first shown by G. Wannier\,\citep{Wannier1962}
that this model has an exact solution consisting of a \textit{Wannier-Stark
ladder} spectrum made up of a discrete set of non-degenerate and equally
spaced energy levels, $\varepsilon_{m}=\!maeE$, with $m\!\in\!\mathbb{Z}$.\,\,The
corresponding eigenstates are the so-called Wannier-Stark states (WSSs)
which, in a real-space representation, are given by\,\citep{Fukuyama1973,Holthaus1996}

\begin{equation}
\ket{\Psi_{m}}=\sum_{n=-\infty}^{\infty}\psi_{m}(n)\ket n=\sum_{n=-\infty}^{\infty}J_{n-m}\left(\frac{2w}{aeE}\right)\ket n,\label{eq:WSS}
\end{equation}
where $J_{n}(x)$ are Bessel functions of the first-kind. State $\ket{\Psi_{m}}$
is centered at the site $m$ and has an effective half-width of $\ell_{{\scriptscriptstyle \text{WS}}}=2w/(aeE)$.
For $\abs{n-m}\gg\ell_{{\scriptscriptstyle \text{WS}}}$, the WSS
decay exponentially as $\left|\psi_{m}(n)\right|\sim e^{-\left|n-m\right|/\xi_{\text{WS}}}$,
with $\xi_{\text{WS}}^{-1}=\log\left(\ell_{{\scriptscriptstyle \text{WS}}}^{-1}\right)$.
It is important to notice that for strong biases, $aE\gg w$, we have
that $\xi_{{\scriptscriptstyle \text{WS}}}\ll\ell_{{\scriptscriptstyle \textrm{WS}}}$,
meaning that in theses conditions the central eigenstates of a finite
chain are well approximated by the WSSs of the infinite system. In
Appendix\,\ref{sec:Bloch-Oscillations-and}, we review how WSSs give
origin to current BOs. 

Following a partition-free approach to transport \citep{Cini80},
for times $t<0$, the leads and central region are connected and in
thermodynamic equilibrium. The initial state is thus characterized
by the reduced density matrix
\begin{equation}
\rho_{0}=f\left(\mathcal{H}_{0}\right)=\sum_{\alpha}f\left(\varepsilon_{0,\alpha}\right)\left|\psi_{0,\alpha}\right\rangle \left\langle \psi_{0,\alpha}\right|\label{eq:initial_state}
\end{equation}
where $\varepsilon_{0,\alpha}$ and $\left|\psi_{0,\alpha}\right\rangle $
are the eigenenergies and eigenstates of the initial Hamiltonian $\mathcal{H}_{0}\equiv\mathcal{H}(t<0)$.
$f(\epsilon)=\left[e^{\beta\left(\epsilon-\mu\right)}+1\right]^{-1}$
is the Fermi distribution function, with $\beta^{-1}=k_{B}T$ the
inverse temperature and $\mu$ the common Fermi energy. For concreteness,
we will assume that the system is initially at half-filling, $\mu=0$,
and restrict ourselves to the zero temperature limit. At $t=0$, the
electric field is switched on, driving the system away from equilibrium
and generating current flow.

\subsection{\label{subsec:Method-of-Quantum}Method of Quantum Time-Evolution}

Our numerical study simulates the time-dependent charge current that
traverses a bond in the system, once the electric field in the central
region has been turned on. The local current going from site $n\!\to\!n\!+\!1$
is represented by the operator

\begin{equation}
\mathcal{I}_{{\scriptscriptstyle n,n+1}}=-i\left(\ket{n+1}\bra n-\ket n\bra{n+1}\right),\label{eq:CurrentOper}
\end{equation}
whose time-dependent expectation value is given by

\begin{align}
I_{{\scriptscriptstyle n,n+1}}(t) & =\text{Tr}\left[\rho_{0}e^{i\mathcal{H}_{{\scriptscriptstyle +}}t}\mathcal{\mathcal{I}}_{{\scriptscriptstyle n,n+1}}e^{-i\mathcal{H}_{{\scriptscriptstyle +}}t}\right]\nonumber \\
 & =2\text{Im}\bra ne^{i\mathcal{H}_{{\scriptscriptstyle +}}t}\rho_{0}e^{-i\mathcal{H}_{{\scriptscriptstyle +}}t}\ket{n+1},\label{eq:LocalCurrent}
\end{align}
where $\rho_{0}$ is the initial reduced density matrix (\ref{eq:initial_state})
for the partition-free setup, and $\mathcal{H}_{{\scriptscriptstyle +}}\equiv\mathcal{H}(t>0)$
is the Hamiltonian after the electric field is turned on, which is
constant for $t>0$. By defining 
\begin{align}
\ket{\Psi_{t}^{n+1}} & =e^{-i\mathcal{H}_{+}t}\ket{n+1},\\
\ket{\Psi_{t}^{n}} & =e^{-i\mathcal{H}_{+}t}\ket n,\\
\ket{\chi_{t}^{n}} & =f\left(\mathcal{H}_{0}\right)\ket{\Psi_{t}^{n}},\label{eq:chi_t_state}
\end{align}
the expectation value of the current can be written as an inner product
\begin{equation}
I_{{\scriptscriptstyle n,n+1}}(t)=2\text{Im}\left\langle \chi_{t}^{n}\left|\Psi_{t}^{n+1}\right.\right\rangle .
\end{equation}

\begin{figure}[t]
\centering{}\includegraphics[width=1\columnwidth]{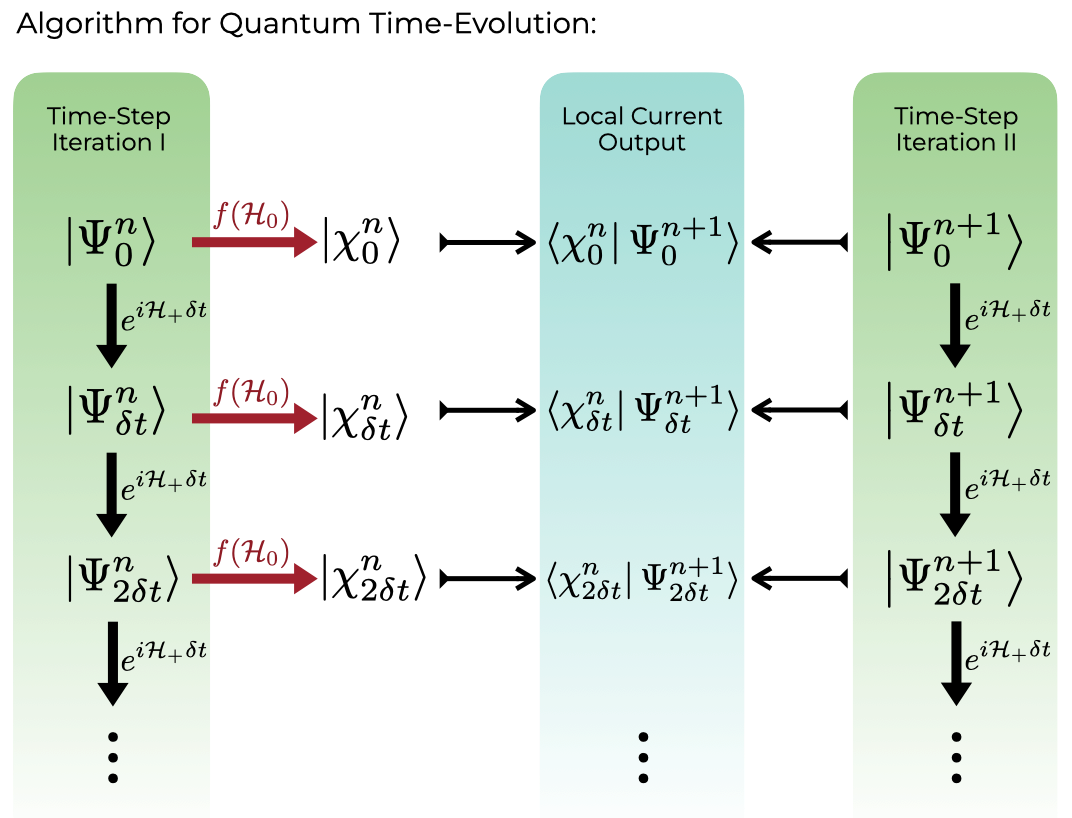}\caption{\label{fig:Algorithm}Scheme of the algorithm used to time-evolve
the system of Fig.\,\ref{fig:models}.}
\end{figure}
In order to evaluate the current of the system, we will consider finite
leads, such that the whole system (central region + leads) has $N$
sites. As shown by Santos Pires \textit{et al}.\,\citep{Pires20},
truncation of the leads does not affect the current for times $t<L_{l}/w_{l}$,
where $L_{l}$ is the number of sizes of the lead, after which effects
of electron reflection at the chain boundaries start to manifest.
Truncation of the system allows for a simple expansion of the time
evolution operator, $e^{i\mathcal{H}_{{\scriptscriptstyle +}}t}$,
and initial reduced density matrix, $\rho_{0}=f\left(\mathcal{H}_{0}\right)$,
in terms of Chebyshev polynomials\citep{Weisse2006} of $\mathcal{H}_{+}$
and $\mathcal{H}_{0}$, respectively. Explicitly we have

\begin{align}
\exp\left[-i\mathcal{H}_{{\scriptscriptstyle +}}t\right] & \approx\sum_{m=0}^{M_{t}}\frac{2(-i)^{m}}{1+\delta_{m,0}}J_{m}\left(\Delta_{\epsilon}t\right)T_{m}\left(\frac{\mathcal{H}_{{\scriptscriptstyle +}}}{\Delta_{\epsilon}}\right),\label{eq:TimeEvolutionExpansion}\\
f\left(\mathcal{H}_{0}\right) & \approx\sum_{m=0}^{M_{\rho}}\frac{2\mu_{m}^{\rho}}{1+\delta_{m,0}}T_{m}\left(\frac{\mathcal{H}_{0}}{\Delta_{\epsilon}}\right),\label{eq:FermiDiracExpansion}
\end{align}
where $\Delta_{\epsilon}$ is a positive energy scale that normalizes
the Hamiltonian spectrum to be within $[-1,1]$, $J_{m}(x)$ is a
Bessel function of the first kind, $T_{m}(x)$ is a Chebyshev polynomial
of the first kind, and $M_{\rho}$ /$M_{t}$ indicate the truncation
order of each expansion. While the form of the expansion coefficients
for the time-evolution operator are know analytically \citep{Tal-Ezer1984},
the values of $\mu_{m}^{\rho}$ must be determined by evaluating the
integral

\begin{equation}
\mu_{m}^{\rho}=\int_{-1}^{1}du\frac{T_{m}(u)}{\pi\sqrt{1-u^{2}}\left[1+e^{\beta\left(\Delta_{\epsilon}u-\mu\right)}\right]},
\end{equation}
which can be easily done numerically. For times $t>0$, we evaluate
the current at discrete mesh of $N_{t}$ points --- $\{0,\delta t,2\delta t,\cdots,t_{\text{max}}\}$
--- with a time step of $\delta t=t_{\text{max}}/N_{t}$. Therefore,
we can write the short time evolution, $\ket{\Psi_{k\delta t}^{n}}=e^{-i\mathcal{H}_{+}\delta t}\ket{\Psi_{\left(k-1\right)\delta t}^{n}}$
$k=1,...,N_{t}$, using Eq. (\ref{eq:TimeEvolutionExpansion}). The
application of $f\left(\mathcal{H}_{0}\right)$, in Eq.~(\ref{eq:chi_t_state}),
is implemented using Eq.~(\ref{eq:FermiDiracExpansion}). Crucial
for the performance of the method is the fact that it only requires
the evaluation of the action of $e^{-i\mathcal{H}_{+}\delta t}$ and
$f\left(\mathcal{H}_{0}\right)$ on states $\ket{\Psi_{k\delta t}^{n}}$.
When doing so, quantities of the form $\ket{\Psi_{k\delta t}^{n}(m)}\equiv T_{m}\left(\mathcal{M}\right)\ket{\Psi_{k\delta t}^{n}}$,
with $\mathcal{M}=\mathcal{H}_{+/0}/\Delta_{\epsilon}$, can be efficiently
evaluated using the Chebyshev recursion
\begin{equation}
\ket{\Psi_{k\delta t}^{n}(m+2)}=2\mathcal{M}\ket{\Psi_{k\delta t}^{n}(m+1)}-\ket{\Psi_{k\delta t}^{n}(m)},
\end{equation}
starting with $\ket{\Psi_{k\delta t}^{n}(0)}=\ket{\Psi_{k\delta t}^{n}}$
and $\ket{\Psi_{k\delta t}^{n}(1)}=\mathcal{M}\ket{\Psi_{k\delta t}^{n}}$.
Therefore, the method only requires matrix-vector multiplications
and has a computational complexity of $\mathcal{O}\!\left(N_{t}\,N\,M_{t}\,M_{\rho}\right)$,
for sparse Hamiltonians. The implementation scheme is illustrated
in Fig.~(\ref{fig:Algorithm}). 

\begin{figure}[t]
\begin{raggedright}
\includegraphics[width=1\columnwidth]{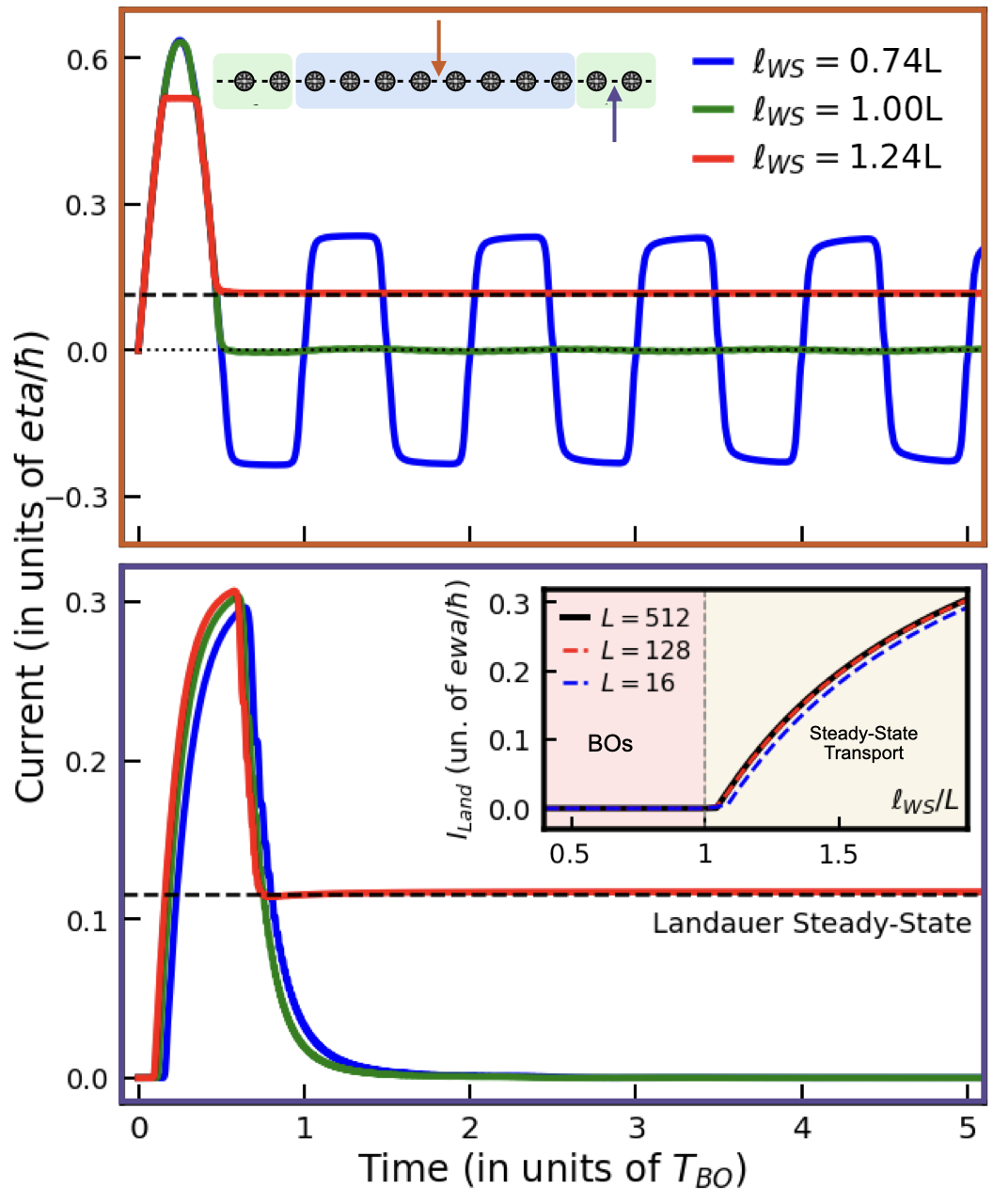}
\par\end{raggedright}
\caption{\label{fig:timeevolveBOS}Current in mesoscopic device as a function
of time (measured in units of $T_{\text{BO}}=2\pi/(eaE)$) measured
inside the central region (top panel), and measured in the right lead
(bottom panel), for different values of the electric field in the
central region. The dashed lines represent the value of the steady
state current, for the cases in which it exists. Inset of bottom panel:
Landauer steady-state current as a function of the Wannier-Stark localization
length divided by the size of the sample, for various sizes. The vertical
dashed line marks the limiting value $\ell_{{\scriptscriptstyle \textrm{WS}}}=L$,
beyond which a non-zero steady-state current emerges. A central region
with 257 sites ($L=128$), leads with $L_{l}=16000$ sites and $w_{l}=w$
was used. }
\end{figure}

\section{Bloch Oscillations Within a Mesoscopic Device\label{sec:From-Infinite-to}}

\begin{figure}[t]
\begin{centering}
\includegraphics[width=1\columnwidth]{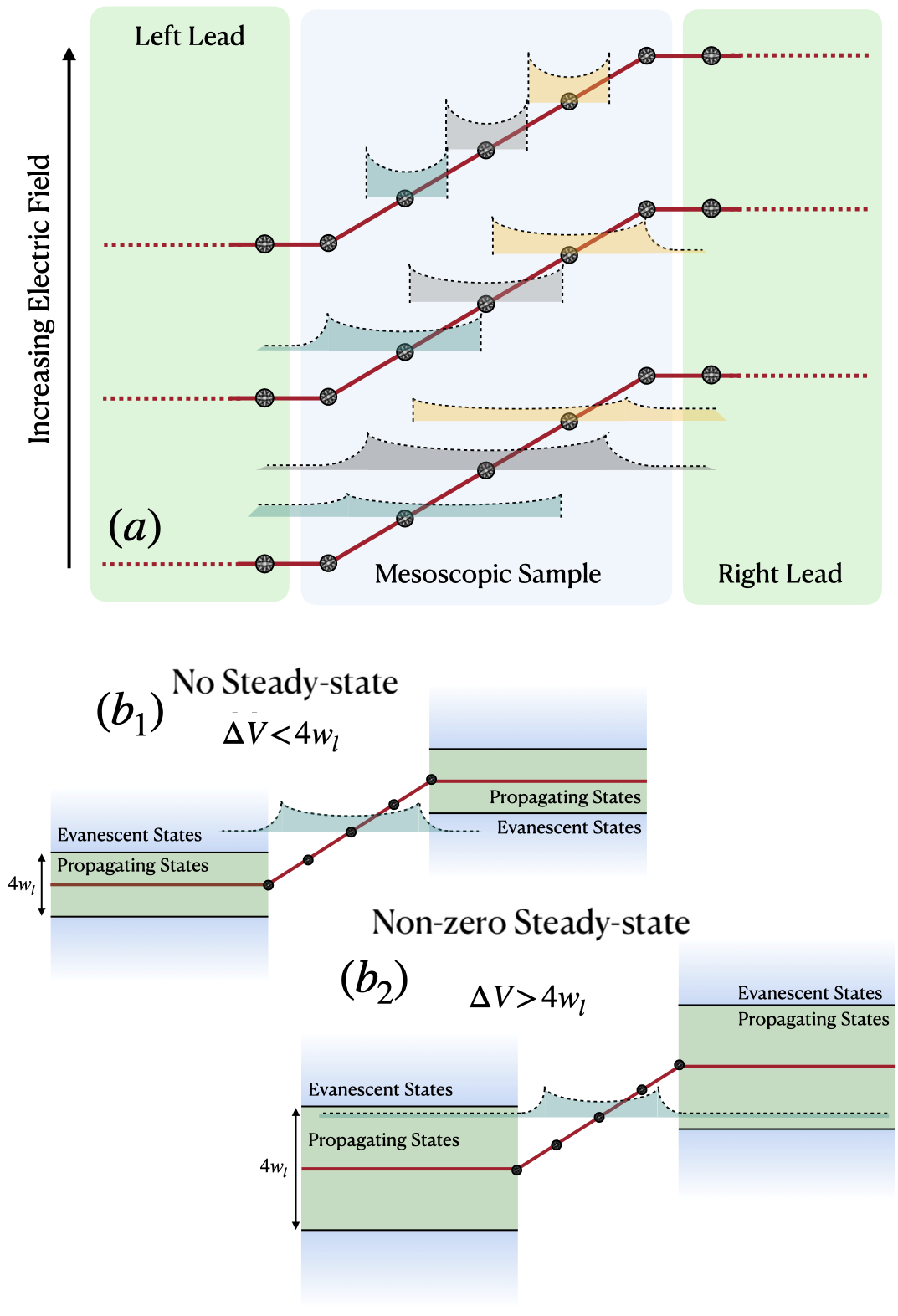}
\par\end{centering}
\caption{\label{fig:DelocalizationWannierStark}Visualization of the conditions
for (a) observation of Bloch oscillations, (b) formation of a non-zero
steady state current. }
\end{figure}

We will start by studying the case when the hopping in the leads and
the central region are the same, $w_{l}=w$, a case which was previously
discussed by Popescu and Croy\,\citep{Popescu2017}. We show the
evaluated current for different values of the electric field, measured
inside the central region, in the top panel of Fig.~\ref{fig:timeevolveBOS}.
For large values of the electric field (small values of $\ell_{\text{WS}}$)
the current displays an oscillatory behaviour with period $T_{\text{BO}}=2\pi/(eaE)$.
As the electric field is reduced, the oscillations become deformed
(clipped) in time. Finally, we observe that as the electric field
is reduced such that $\ell_{\text{WS}}>L$, the Bloch oscillations
disappear and the current tends to a constant value. The current measured
in the leads is zero when Bloch oscillations are observed in the central
region, as shown in the bottom panel of Fig.~\ref{fig:timeevolveBOS}.
When Bloch oscillations are absent, a steady state develops and the
current in the leads tends to the same constant value as the current
inside the central region. The condition for the observation of BOs,
$\ell_{\text{WS}}<L$, can be interpreted in terms of the localization
properties of the WSSs. For strong electric field, the eigenstates
of $\mathcal{H}_{+}$ will be nearly indistinguishable from the WSSs
of an infinite Wannier-Stark chain. Since the current is a local operator,
we expect that its expected value will then have the same oscillations
as the ones of a Wannier-Stark chain. As the electric field is reduced,
eigenstates localized at the center of the sample will remain largely
unchanged, but the states closer to the edges of the central region
will start to leak into the closest lead. As such, these states will
not contribute to the Bloch oscillations, which will thus become clipped.
Finally, if the bias becomes too small, the most central state of
the system will eventually become delocalized, bridging the two leads
and carrying a steady-state current. This mechanism is illustrated
in Fig.~\ref{fig:DelocalizationWannierStark}(a). Notice that the
previous argument does not tell us anything about the value of the
current in leads. In particular, it provides no explanation why the
current is zero there when BOs occur. In order to do so, we must analyse
the spectrum of the leads. The Landauer formula tells us that to obtain
a steady-state current we must have an electron in an occupied state
of one lead tunneling into an empty state of the other lead. Therefore,
the spectrum of the leads must overlap in energy. A one dimensional
tight-binding model with nearest-neighbour hopping $w_{l}$ has a
spectrum with a bandwidth of $4w_{l}$. If the leads are half-filled,
the spectra of the left and right leads overlap provided $\Delta V<4w_{l}$,
as depicted in Fig.~\ref{fig:DelocalizationWannierStark}(b) and
a non-zero steady-state current is possible. If $\Delta V>4w_{l}$,
since there is no overlap between the spectra of the two leads, there
is no propagating state that connects both leads and the steady-state
current must be zero. In this case, incoming electrons from one lead
suffer total reflection as the other lead does not support propagating
states at that energy. If $w_{l}=w$, and recalling $\ell_{\text{WS}}=2w/\left(eEa\right)$,
we have that the condition for the observation of BOs, $\ell_{\text{WS}}<L$,
coincides with the condition for zero steady-state current, $\Delta V>4w$,
a condition previously found by Popescu and Croy\,\citep{Popescu2017}.
This is in agreement with the results for the steady-state current
in Fig.\,\ref{fig:timeevolveBOS}, for different central sample sizes
and different values of $\ell_{{\scriptscriptstyle \textrm{WS}}}$,
obtained using the Landauer formula as implemented in the Kwant package\,\citep{Groth2014}.

\begin{figure}[t]
\begin{centering}
\includegraphics[width=1\columnwidth]{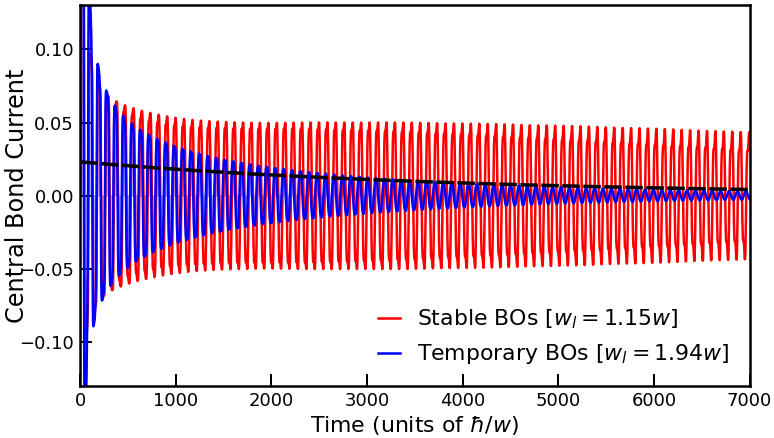}
\par\end{centering}
\caption{\label{fig:DecayingOscilations}Plots of the local electric current
measured over time in the central bond of the mesoscopic sample composed
of 69 sites ($L\!=\!34$) and a bias potential $\Delta V\!=\!5w$.
We showcase two examples of: \textit{(i)} a mesoscopic device supporting
stable clipped BOs if $w_{l}\!=\!1.15w$ ($w_{l}<\Delta V/4$), and
\textit{(ii)} a mesoscopic device having $w_{l}\!=\!1.94w$ ($w_{l}>\Delta V/4$)
that now supports tBOs that decay exponentially in time. Such decay
is seen to correspond to a decaying time given by $\left(1/\tau_{0}+1/\tau_{1}\right)^{-1}$
(dashed black line) for long times.}
\end{figure}

The previous discussion makes clear that the simultaneous observation
of BOs in the central region and zero steady-state current in leads
is an artifact of having the hoppings in the central region and leads
be the same, $w_{l}=w$. Otherwise, the condition for the observation
of BOs, $\ell_{\text{WS}}<L/2\Leftrightarrow4w<\Delta V$, and the
condition for observation of a non-zero steady state current, $\Delta V<4w_{l}$,
become distinct. Therefore, if we are in a regime where $4w<\Delta V<4w_{l}$,
we can expect to simultaneously observe BOs and a steady-state non-zero
current. Indeed, this is what occurs as can be seen in Fig.~\ref{fig:DecayingOscilations},
where we show the current for a case where $w_{l}=w$, with persistent
BOs (a small modulation of the oscillations can be observed, which
is discussed in Appendix~\ref{apx:BOs_modulation}) and another with
$w_{l}\neq w$, for which BOs adquire a finite lifetime and coexist
with a non-zero steady-state current. As an oscillating current precludes
the formation of a steady-state, BOs that coexist with a steady-state
cannot be persistent and must instead be a transient phenomena with
a characteristic decay rate, which we will refer to as transient Bloch
Oscillations (tBOs).

\section{Quasiparticle approximation to transient Bloch Oscillations\label{sec:Relating-the-Lorentzian's}}

\subsection{Quasiparticle states of the central region}

Having established the possibility of transient BOs, we will now develop
approximate theory to describe their decay times. Our starting point
is the Caroli formula\citep{Caroli71}, which expresses the transmittance
$\mathcal{T}(\epsilon)$, at an energy $\epsilon$, in terms of Green's
function as

\begin{equation}
\mathcal{T}(\epsilon)=\text{Tr}\left[\mathbf{G}^{\dagger}(\epsilon)\cdot\boldsymbol{\Gamma}_{\text{R}}(\epsilon)\cdot\mathbf{G}(\epsilon)\cdot\boldsymbol{\Gamma}_{\text{L}}(\epsilon)\right],\label{eq:caroli}
\end{equation}
where is a trace over the central sample's Hilbert space, $\mathbf{G}(\varepsilon)$
is the retarded Green's function of the central sample when connected
to the leads
\begin{equation}
\mathbf{G}(\epsilon)=\left[\epsilon-\mathcal{\boldsymbol{H}}_{{\scriptscriptstyle \textrm{C}}}-\boldsymbol{\Sigma}_{\text{R}}(\epsilon)-\boldsymbol{\Sigma}_{\text{L}}(\epsilon)\right]^{-1},\label{eq:green_func_eq}
\end{equation}
in terms of the isolated central sample's Hamiltonian $\mathcal{\boldsymbol{H}}_{{\scriptscriptstyle \textrm{C}}}$
{[}Eq.~(\ref{eq:Hamiltonian_Central_region}) for $t>0${]} and the
self-energies introduced by the connected semi-infinite leads, $\boldsymbol{\Sigma}_{\text{R/L }}(\epsilon)$,
and the quantities $\boldsymbol{\Gamma}_{\text{R/L}}(\epsilon)=i\left[\boldsymbol{\Sigma}_{\text{R/L }}(\epsilon)-\boldsymbol{\Sigma}_{\text{R/L }}^{\dagger}(\epsilon)\right]$,
are the \textit{level-width matrices}. For semi-infinite tight-binding
chains, one can analytically determine the self-energy, thus arriving
at the expressions\,\citep{Datta1995,Pires20}, 

\begin{equation}
\boldsymbol{\Sigma}_{\text{R/L}}(\epsilon)=w_{l}\Sigma\left(\frac{\epsilon\pm\Delta V/2}{2w_{l}}\right)\ket{\pm L}\bra{\pm L}\label{eq:SelfEnergy}
\end{equation}
where $\Sigma\left(\varepsilon\right)=\varepsilon-i\sqrt{1-\left(\varepsilon+i0^{+}\right)^{2}}$.
The Green's function of the central region $\mathbf{G}(\epsilon)$
can be expressed in terms of its right, $\ket{\Phi_{n}^{R}(\epsilon)}$,
and left, $\bra{\Phi_{n}^{L}(\epsilon)}$, eigenvectors of the effective
(non-hermitian) Hamiltonian of the central region connect to the leads,
$\boldsymbol{\mathcal{H}}_{\text{eff}}(\epsilon)=\mathcal{\boldsymbol{H}}_{{\scriptscriptstyle \textrm{C}}}+\boldsymbol{\Sigma}^{\text{R}}(\epsilon)+\boldsymbol{\Sigma}^{\text{L}}(\epsilon)$.\footnote{\noindent We define the left and right eigenvectors of $\boldsymbol{\mathcal{H}}_{\text{eff}}(\epsilon)=\mathcal{\boldsymbol{H}}_{{\scriptscriptstyle \textrm{C}}}+\boldsymbol{\Sigma}^{\text{R}}(\epsilon)+\boldsymbol{\Sigma}^{\text{L}}(\epsilon)$
as $\text{\ensuremath{\boldsymbol{\mathcal{H}}_{\text{eff}}}(\ensuremath{\epsilon})\ensuremath{\left|\Phi_{n}^{R}(\epsilon)\right\rangle }=\ensuremath{\lambda_{n}(\epsilon)}\ensuremath{\left|\Phi_{n}^{R}(\epsilon)\right\rangle }}$
and $\bigl\langle\Phi_{n}^{L}(\epsilon)\bigr|\boldsymbol{\mathcal{H}}_{\text{eff}}(\epsilon)=\bigl\langle\Phi_{n}^{L}(\epsilon)\bigr|\lambda_{n}(\epsilon)$,
with the eigenvalue written as $\lambda_{n}(\epsilon)=\epsilon_{n}(\epsilon)-i\gamma_{n}(\epsilon)$
. We have that $\bigl\langle\Phi_{n}^{L}(\epsilon)\bigr|$ form a
dual basis to $\bigl|\Phi_{n}^{R}(\epsilon)\bigr\rangle$, $\bigl\langle\Phi_{n}^{L}(\epsilon)\left|\Phi_{m}^{R}(\epsilon)\right.\bigr\rangle=\delta_{n,m}$.
However, since $\boldsymbol{\mathcal{H}}_{\text{eff}}(\epsilon)$
is non-hermitian, we have that $\bigl|\Phi_{n}^{R}(\epsilon)\bigr\rangle\neq\left[\bigl\langle\Phi_{n}^{L}(\epsilon)\bigr|\right]^{\dagger}$.
Notice that $\left[\bigl\langle\Phi_{n}^{L}(\epsilon)\bigr|\right]^{\dagger}\equiv\bigl|\Phi_{n}^{L}(\epsilon)\bigr\rangle$
are the right eigenstates of $\boldsymbol{\mathcal{H}}_{\text{eff}}^{\dagger}(\epsilon)$.}We have that

\begin{equation}
\mathbf{G}(\epsilon)=\sum_{n}\frac{\ket{\Phi_{n}^{R}(\epsilon)}\bra{\Phi_{n}^{L}(\epsilon)}}{\epsilon-\epsilon_{n}(\epsilon)+i\gamma_{n}(\epsilon)},\label{eq:SPGF_Connected}
\end{equation}
where the summation is over the entire Hilbert space of the central
sample, and $\epsilon_{n}(\epsilon)-i\gamma_{n}(\epsilon)$ are the
eigenvalues of $\boldsymbol{\mathcal{H}}_{\text{eff}}(\epsilon)$,
separated into their real and imaginary parts. Notice that both the
eigenvectors, $\ket{\Phi_{n}^{R}(\epsilon)}$ and $\bra{\Phi_{n}^{L}(\epsilon)}$,
and the eigenvalues, $\epsilon_{n}(\epsilon)-i\gamma_{n}(\epsilon)$,
are a function of the energy $\epsilon$. 

\begin{figure}[t]
\includegraphics[width=1\columnwidth]{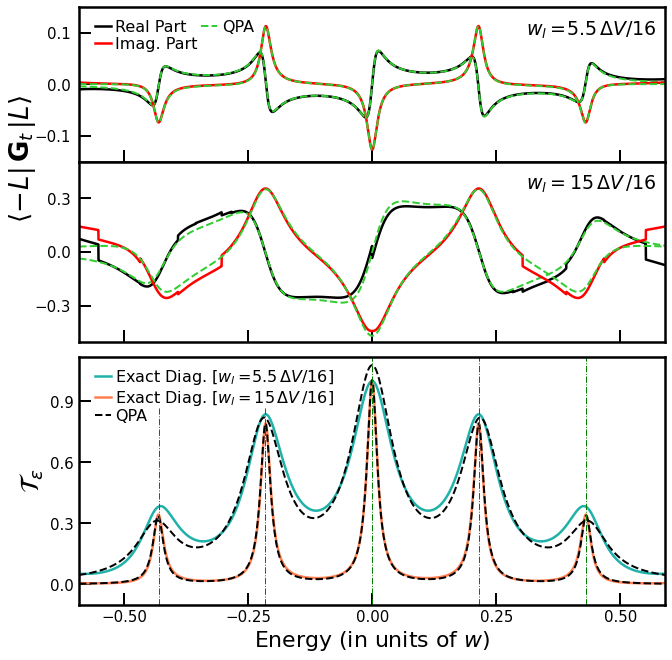}

\caption{\label{fig:QPA_GF_transmittance}Top panel: real and imaginary parts
of the the Green's function $\protect\bra{-L}\mathbf{G}(\epsilon)\protect\ket L$
as a function of energy for a system with $L=12$, $\Delta V=4.3w$
and two values of $w_{l}$. The full lines represent exact results
and the dashed lines represent results obtained within the QPA. Bottom
panel: transmittance $\mathcal{T}(\epsilon)$ as a function of energy
, for the same system as in the top panel. The vertical lines, represent
the energies of the inner most WSS, $\epsilon_{m}=eEam$, $m=0,\pm1,\pm2$. }
\end{figure}

If the states of the central region are only weakly perturbed by the
hybridization with the leads, the eigenvalues, $\ket{\Phi_{n}^{R/L}(\epsilon)}$,
and eigenstates, $\epsilon_{n}(\epsilon)-i\gamma_{n}(\epsilon)$,
will be weakly dependent on the energy $\epsilon$. Furthermore, if
the electric field is strong enough, $\ell_{\text{WS}}\ll L$, the
eigenstates and eigenvalues of $\boldsymbol{\mathcal{H}}_{\text{eff}}(\epsilon)$
will be well approximated by WSS. We will refer to these approximations
as the \textit{weak coupling and strong field approximation}. With
these considerations, we can employ a quasiparticle approximation
(QPA) to the Green's function, $\mathbf{G}(\epsilon)$, in which we
approximate $\ket{\Phi_{n}^{R/L}(\epsilon)}\simeq\ket{\Phi_{n}^{R/L}(\epsilon_{{\scriptscriptstyle \text{C}},n})}\equiv\left|\Phi_{n,\text{QPA}}^{R/L}\right\rangle $
and $\epsilon_{n}(\epsilon)-i\gamma_{n}(\epsilon)\simeq\epsilon_{n}(\epsilon_{\text{C},n})-i\gamma_{n}(\epsilon_{\text{C},n})\equiv\epsilon_{n}^{\text{QPA}}-i\gamma_{n}^{\text{QPA}}$,
where $\epsilon_{\text{C},n}$ are the eigenstates of the isolated
central region, $\mathcal{\boldsymbol{H}}_{\text{C}}$\footnote{In practice, to obtain the QPA to $\mathbf{G}(\epsilon)$ we proceed
as follows. (i) For each energy of the isolated central region, $\epsilon_{\text{C},n}$,
we start by computing the left/righ eigenstates and eigenvalues of
$\boldsymbol{\mathcal{H}}_{\text{eff}}(\epsilon_{\text{C},n})$. (ii)
Then we select the eigenpair $\bigl|\Phi_{m}^{R/L}(\epsilon_{{\scriptscriptstyle \text{C}},n})\bigr\rangle$,
$\epsilon_{m}(\epsilon_{\text{C},n})-i\gamma_{m}(\epsilon_{\text{C},n})$
with $\epsilon_{m}(\epsilon_{\text{C},n})$ closest to $\epsilon_{\text{C},n}$.
(iii) Sum over the contributions obtained in this way for each eigenenergy
of the isolated central region $\epsilon_{\text{C},n}$.}. Therefore, $\epsilon_{n}(\epsilon_{{\scriptscriptstyle \text{C}},n})$
are the corrected energy levels and $\gamma_{n}(\epsilon_{{\scriptscriptstyle \text{C}},n})$
the corresponding broadenings or decay rates induced by the hybridization
of the central region with the leads. Within the QPA we have that
\begin{equation}
\mathbf{G}(\epsilon)\simeq\mathbf{G}_{\text{QPA}}(\epsilon)=\sum_{n}\frac{\ket{\Phi_{n,\text{QPA}}^{R}}\bra{\Phi_{n,\text{QPA}}^{L}}}{\epsilon-\epsilon_{n}^{\text{QPA}}+i\gamma_{n}^{\text{QPA}}}.\label{eq:QPA_GF}
\end{equation}
Within the weak coupling and strong field approximations, we also
have that $\left|\epsilon_{n}^{\text{QPA}}-\epsilon_{n+1}^{\text{QPA}}\right|\gtrsim\gamma_{n}^{\text{QPA}},\gamma_{n+1}^{\text{QPA}}$,
which allows us to further approximate the transmission, which can
be written as $\mathcal{T}(\epsilon)=\Gamma_{\text{L}}(\epsilon)\Gamma_{\text{R}}(\epsilon)\left|\bra{-L}\mathbf{G}(\epsilon)\ket L\right|^{2}$,
as a sum of Lorentzians
\begin{multline}
\mathcal{T}(\epsilon)\simeq\mathcal{T}_{\text{QPA}}(\epsilon)=\\
=\Gamma_{\text{L}}(\epsilon)\Gamma_{\text{R}}(\epsilon)\sum_{n}\frac{\left|\braket L{\Phi_{n,\text{QPA}}^{R}}\right|^{2}\left|\braket{\Phi_{n,\text{QPA}}^{L}}{-L}\right|^{2}}{\left(\epsilon-\epsilon_{n}^{\text{QPA}}\right)^{2}+\left(\gamma_{n}^{\text{QPA}}\right)^{2}},\label{eq:ApproximateTransmittance}
\end{multline}
where $\Gamma_{\text{L/R}}(\epsilon)=\sqrt{\left(2w_{l}\right)^{2}-\left(\epsilon\pm\Delta V/2\right)^{2}}$.
In Fig.~\ref{fig:QPA_GF_transmittance}, we show the real and imaginary
parts of $\bra{-L}\mathbf{G}(\epsilon)\ket L$ and of the transmission
computed both exactly and within the QPA. The exact results for the
transmittance were obtained using the $\texttt{Kwant}$ package\,\citep{Groth2014}.
We can see that the QPA works remarkably well, provided we are in
the conditions for strong field and weak coupling, $(\ell_{\text{WS}}\ll L$),
for the most central states of the device (which are the least hybridized
with the leads). We can see that the transmittance indeed approaches
a sum of Lorentzian functions centered at the QPA energies $\epsilon_{n}^{\text{QPA}}$
and with width given by $\gamma_{n}^{\text{QPA}}$. As we will see
in the next subsection, the decay rate of the tBOs are related to
$\gamma_{n}^{\text{QPA}}$. Interestingly, we see the that the width
of the Lorentzians reduces with and increasing $w_{l}$. 

\subsection{Transient current due to quasiparticle states}

\begin{figure}
\includegraphics[width=1\columnwidth]{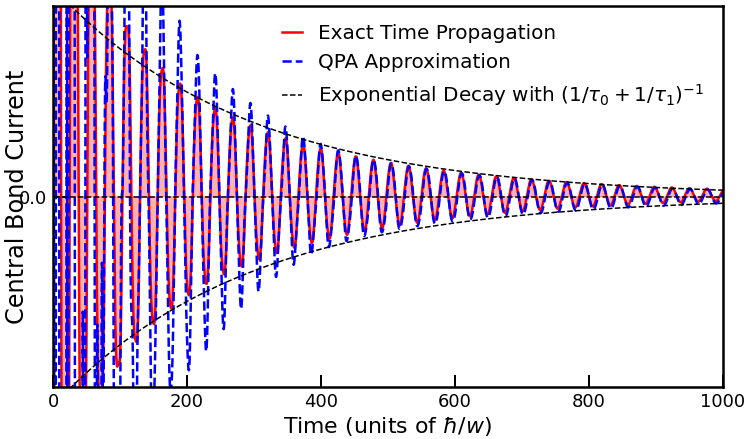}

\caption{\label{fig:QPA_transient_current}Current measured in the central
region computed exactly (solid, red line) and within the QPA (dashed,
blue line) {[}Eq.~(\ref{eq:transient_current_QPA}){]}. }
\end{figure}

We will now develop a time-resolved theory for tBOs based on the QPA.
We will start by arguing that the tBO are a phenomena that depends
on the local properties of the central region that is subjected to
the electric field. This is illustrated by the results of Fig.~\ref{fig:timeevolveBOS},
where we can see that BOs can be observed in the current measured
in the central region, but not on the leads. Furthermore, we know
that the steady-steady current depends only on the occupation of the
leads, with effects due to the occupation of the central region being
washed-out. Therefore, this further reinforces the notion that tBO
depend mostly on the occupation of states in the central region. Therefore,
we approximate the current measured in the central region as
\begin{equation}
I_{n,n+1}(t)\simeq I_{\text{Land}}+I_{n,n+1}^{\text{Trans}}\left(t\right),
\end{equation}
where $I_{\text{Land}}$ is the Landauer steady-state current, which
is controlled by the occupation of the leads, and 
\begin{equation}
I_{n,n+1}^{\text{Trans}}\left(t\right)=\hbar^{2}\text{Tr}\left[\rho_{\text{C},0}\mathbf{G}^{\dagger}(t)\mathcal{I}_{n,n+1}\mathbf{G}(t)\right]
\end{equation}
approximates the current due to the occupation of the central region,
which will capture the tBO. In the above expression $\rho_{\text{C},0}$
is the projection of the initial (partition-free) reduced density
matrix onto the central region, and $\mathbf{G}(t)$ is the projection
of the full time evolution operator $e^{-\frac{i}{\hbar}\mathcal{H}_{+}t}$
into the central region, which is nothing more than the retarded Green's
function of the central region. Within the QPA, we use Eq.~(\ref{eq:QPA_GF}),
which leads to
\begin{multline}
\mathbf{G}(t)=\int\frac{d\epsilon}{2\pi\hbar}e^{-i\epsilon t/\hbar}\mathbf{G}(\epsilon)\\
\simeq-\frac{i}{\hbar}\Theta(t)\sum_{n}e^{-i\left(\epsilon_{n}^{\text{QPA}}-i\gamma_{n}^{\text{QPA}}\right)t/\hbar}\ket{\Phi_{n,\text{QPA}}^{R}}\bra{\Phi_{n,\text{QPA}}^{L}}.
\end{multline}
We therefore, obtain the approximate equation for the transient current
inside the central region
\begin{multline}
I_{n,n+1}^{\text{Trans}}\left(t\right)\simeq\sum_{m,r}e^{-\left(\gamma_{m}^{\text{QPA}}+\gamma_{r}^{\text{QPA}}\right)t/\hbar}e^{-i\left(\epsilon_{m}^{\text{QPA}}-\epsilon_{r}^{\text{QPA}}\right)t/\hbar}\times\\
\times\bra{\Phi_{m,\text{QPA}}^{L}}\rho_{\text{C},0}\ket{\Phi_{r,\text{QPA}}^{L}}\bra{\Phi_{r,\text{QPA}}^{R}}\mathcal{I}_{n,n+1}\ket{\Phi_{m,\text{QPA}}^{R}}.\label{eq:transient_current_QPA}
\end{multline}
In Fig.~(\ref{fig:QPA_transient_current}), we compare the exact
result of the time-resolved current with the estimation of the transient
current within the QPA. We see that for large enough times we obtain
an excellent agreement. For shorter times we see significant differences.
We attribute these differences to the fact that at short time scales,
the current will be dominated by states strongly hybridized with the
leads, for which the QPA fails. For large enough times, the decay
of the tBOs is well approximated by a single exponential. To estimate
its effective decay time, we first notice that deep inside the central
region, the quasiparticle eigenstates are well approximated by the
WSS, which are purely real, which allow us to conclude that $\bra{\Phi_{m,\text{QPA}}^{R}}\mathcal{I}_{n,n+1}\ket{\Phi_{m,\text{QPA}}^{R}}\simeq0$.
In addition, we also expect the inner most states of the central region
will have the smallest decay rates, as these are more weakly coupled
to the leads. Finally, the diagonal contributions to the transient
current in equation \ref{eq:transient_current_QPA} will not contribute
with an oscillatory dynamic, since the power of the complex exponential
will vanish for these terms. For these reasons, we conclude that for
relatively long times, the sum in Eq.~(\ref{eq:transient_current_QPA})
will be dominated by the contributions from $(m,r)=(0,\pm1),(\pm1,0)$,
which leads to an effective decay time for the tBOs of $\tau_{\text{eff}}=\left(\gamma_{0}^{\text{QPA}}+\gamma_{1}^{\text{QPA}}\right)^{-1}$,
where we used the fact that $\gamma_{1}^{\text{QPA}}=\gamma_{-1}^{\text{QPA}}$.
As shown in Figs.~\ref{fig:DecayingOscilations} and \ref{fig:QPA_transient_current},
the decay of the tBO for large times is well captured by $\tau_{\text{eff}}$.

\subsection{Quasianalytic estimation of the tBO decay time\label{subsec:Quasianalytic-estimation-of}}

\begin{figure}[t]
\begin{centering}
\includegraphics[width=1\columnwidth]{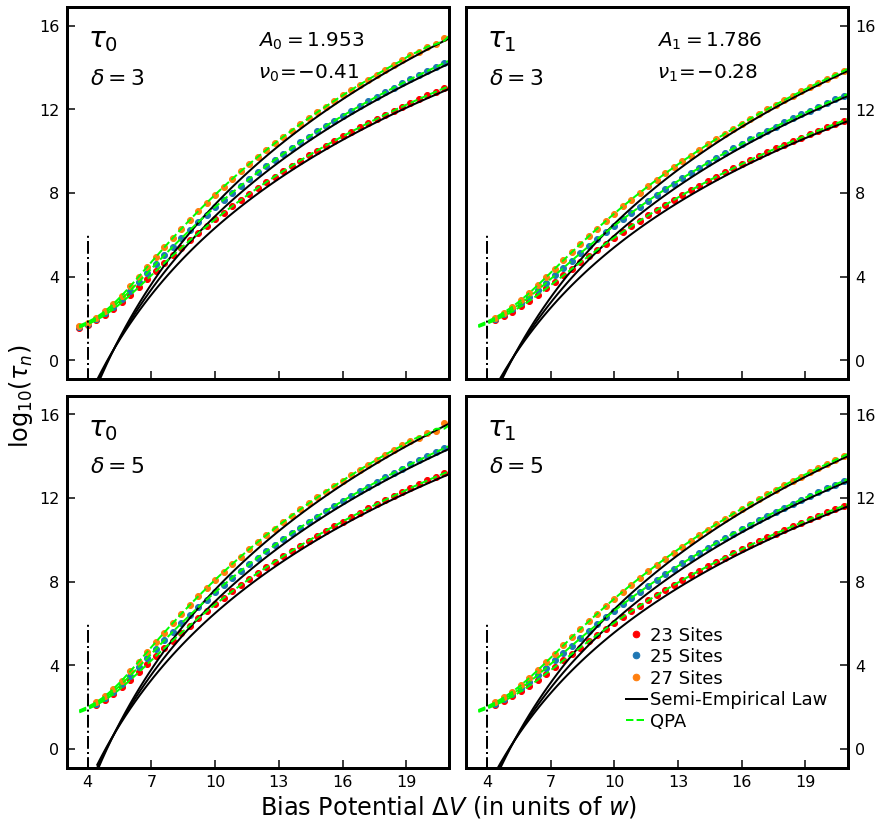}
\par\end{centering}
\centering{}\caption{\label{fig:Lifetimes}Lifetimes of the WSSs centered on the central
($\tau_{0}$) and first off-center sites ($\tau_{1}\!=\!\tau_{-1}$)
of the mesoscopic sample, as a function of the potential bias $\Delta V$.
The dots mark the numerical values extracted from the HWHW of the
central lorentzian in quantum transmittance of the biased central
sample. Two mesoscopic devices are considered, having $\delta\!=\!3$
(upper panels) and $\delta\!=\!5$, with the black lines representing
a fit of the numerical data to the semi-empirical expressions of Eq.\,(\ref{eq:Lifetime}).
The corresponding values of $A_{n}$ and $\nu_{n}$ are shown in the
panels.}
\end{figure}

We will now provide a quasianalytic expression for the decay time
of the inner most quasiparticle states of the central region. To do
so, we start by noticing that within the QPA the decay rates can be
approximated by
\begin{equation}
\gamma_{n}^{\text{QPA}}\simeq-\text{Im}\bra{\Phi_{n,\text{QPA}}^{L}}\boldsymbol{\Sigma}_{\text{R}}(\epsilon_{\text{C},n})+\boldsymbol{\Sigma}_{\text{L}}(\epsilon_{\text{C},n})\ket{\Phi_{n,\text{QPA}}^{R}}.
\end{equation}
In the previous equation, we might be tempted to further approximate
the quasiparticle states by the WSS. However, even though the quasiparticle
eigenstates are well approximate by WSS deep within the central region,
closer to the edges significant differences can be observed. We make
instead the following ansatz
\begin{equation}
\gamma_{n}^{\text{QPA}}\simeq-\frac{1}{C_{n}\left(w_{l},\Delta V\right)}\text{Im}\bra{\Psi_{n}}\boldsymbol{\Sigma}_{\text{R}}(\epsilon_{\text{C},n})+\boldsymbol{\Sigma}_{\text{L}}(\epsilon_{\text{C},n})\ket{\Psi_{n}},
\end{equation}
where $\ket{\Psi_{n}}$ are the WSS {[}Eq.~(\ref{eq:WSS}){]} and
we assumed that $C_{n}\left(w_{l},\Delta V\right)$ is independent
of the central region size. We can now evaluate analytically
\begin{multline}
\text{Im}\bra{\Psi_{n}}\boldsymbol{\Sigma}_{\text{R}}(\epsilon_{\text{C},n})+\boldsymbol{\Sigma}_{\text{L}}(\epsilon_{\text{C},n})\ket{\Psi_{n}}=\\
=\frac{1}{2}\Gamma_{\text{L}}(\epsilon_{\text{C},n})J_{L+n}^{2}\left(\ell_{{\scriptscriptstyle \text{WS}}}\right)+\frac{1}{2}\Gamma_{\text{R}}(\epsilon_{\text{C},n})J_{L-n}^{2}\left(\ell_{{\scriptscriptstyle \text{WS}}}\right).\label{eq:termoperturb}
\end{multline}
To make progress, we recall that the QPA is only valid in the strong
field and weak coupling limit, $\ell_{\text{WS}}\ll L$. Furthermore,
we will focus on the decay rate of the three innermost states, $n=0,\pm1$,
which dominate the decay rate of the tBOs for long enough times. Specializing
to $n=0,\pm1$ in the limit $\ell_{\text{WS}}\to0^{+},$we approximate
$\Gamma_{\text{L}}(\epsilon_{\text{C},n})\simeq\Gamma_{\text{R}}(\epsilon_{\text{C},n})=\sqrt{4w_{l}^{2}-\Delta V^{2}}+\mathcal{O}\left(n/L\right)^{2}$
and approximate the Bessel functions as (see Abramowitz and Stegun\,\citep{abramowitzHandbookMathematicalFunctions1964}):

\begin{equation}
J_{L+n}\left(\ell_{{\scriptscriptstyle \text{WS}}}\right)\simeq\frac{1}{\sqrt{2\pi\left(L+n\right)}}\left(\frac{e\ell_{{\scriptscriptstyle \text{WS}}}}{2\left(L+n\right)}\!\right)^{L+n}.\label{eq:Aprrox}
\end{equation}
We therefore obtain the approximate expression for the decay times

\begin{multline}
\frac{\tau_{n}^{\text{QPA}}}{4\pi}=\frac{1}{4\pi}\frac{1}{\gamma_{n}^{\text{QPA}}}\simeq\\
\simeq\frac{C_{n}\left(\delta,\Delta V\right)\left(L-\abs n\right)}{(2-\delta_{0n})\sqrt{\delta\left(\delta+2\right)}}\left[\frac{L-\abs n}{4e\left(L+1\right)}\!\right]^{{\scriptscriptstyle 2\left(L-\abs n\right)}}\Delta V^{{\scriptscriptstyle 2\left(L-\abs n\right)-1}},\label{eq:Lifetime}
\end{multline}
where we introduced the parameter $\delta=4w_{l}/\Delta V-1$ that
describes the overlap between the bands of propagating states in both
leads. Note that tBOs only exist if $\delta>0$ and, as shown in Eq.\,(\ref{eq:Lifetime}),
the lifetimes associated to the central most WSSs of the mesoscopic
sample diverge as $\delta\to0^{+}$. At the same time, we were able
to extract analytically that $\tau_{n}^{\text{QPA}}$ has a very steep
power-law dependence on the bias potential, with an exponent that
grows linearly with the sample size. Note that this is a consequence
of the exponential tails of the WSSs which, if $\ell_{{\scriptscriptstyle \text{WS}}}\ll L$,
fully determine the way in which they are affected by the presence
of the leads. We assume that $C_{n}\left(\delta,\Delta V\right)=A_{n}(\delta+1)^{2}\Delta V^{\nu_{n}}$,
$A_{n}$ and $\nu_{n}$ depend only on $n$ and can be determined
to fit the numerical data (this assumption is numerically validated
in Appendix (\ref{sec:Study-of-Function})). Note that the dependence
of $\tau_{n}^{\text{QPA}}$ on the parameter $\delta$ is fully fixed,
which is to say that the dependence on $w_{l}$ is completely determined.
Furthermore, the extra power law dependence on $\Delta V$ coming
from $C_{n}\left(\delta,\Delta V\right)$ is much weaker than the
one coming from equations (\ref{eq:termoperturb}) and (\ref{eq:Aprrox}),
as the coefficients $\nu_{n}$ were determined to be 0.41 and 0.28
for $n=0,1$ respectively. We cannot say if the correction coming
from the functions $C_{n}(\delta,\Delta V)$ to the overall dependence
of equation \ref{eq:Lifetime} on $\Delta V$ is trully a power law
or a logarithmic correction, since we have not probed these functions
with a large enough $\Delta V$ interval. Nevertheless, this should
not matter in the limit of large $L$, as this correction is dominated
by the power-law that comes from equation \ref{eq:termoperturb}.
In Fig.\,\ref{fig:Lifetimes} we compare this semi-analytic expression
with results obtained within the fully numeric QPA and those obtained
by fitting Lorentzian functions to the calculated transmission function
for $\tau_{0}$ and $\tau_{1}$.

\subsection{Numerical extraction of decay times}

Having devised a theoretical model to analyze the decaying current
oscillations in the strong field and weak coupling regime, we are
now in position to complete the analysis of the numerical simulation
results first shown in Fig.~\ref{fig:DecayingOscilations}. In Fig.~\ref{fig:FourierAnalysis}(a),
we show results for the time-dependent current crossing the central
bond of a mesoscopic sample ($25$ sites) subject to different potential
biases, $\Delta V$. In all the case, the bandwidths of the leads
was adjusted such that $\delta=4w_{l}/\Delta V-1=0.25$ remains constant,
thus guaranteeing the existence of tBOs in the central sample. Firstly,
we see that in all cases the current displays damped oscillations
that decay towards a constant value after a few periods of oscillation.
By applying the two-terminal Landauer formula, we further conclude
that the asymptotic current corresponds to the Landauer current, $I_{\text{Land}}$,
of each strongly biased sample. Secondly, we also observe that the
decay time of these oscillations (as well as the pseudo-period) increase
with $\Delta V$ as expected from our theoretical understanding of
this phenomenon. In fact, as depicted in the inset of Fig.\,\ref{fig:FourierAnalysis}(a),
a rescaling of the time variable by the corresponding $\tau_{\text{eff}}$
serves to collapse the decaying envelope of all the curves, which
proves that this is the indicated time-scale.

While the previous analysis seemingly demonstrated that our theoretical
model for the tBOs serves to explain the behavior of the current inside
a strongly biased mesoscopic sample, we can perform a more precise
analysis of the current oscillations in Fig.\,\ref{fig:FourierAnalysis}\,(a).
We will focus on the current measured at the center of the sample
between sites 0 and 1, $I_{0,1}\!\left(t\right)$. For that purpose,
we begin by Fourier transforming $I_{0,1}\!\left(t\right)$ into the
frequency-domain ($\omega$-domain) which gives rise to the data points
plotted in Figs.\,\ref{fig:FourierAnalysis}(b). If our model for
the tBOs is accurate, then the local current at the center of the
sample for long enough times, as per equations \ref{eq:LocalCurrent}
and \ref{eq:transient_current_QPA}, is given by

\begin{equation}
I_{0,1}\!\left(t\!\gg\!0\right)\simeq A\cos\left(\Omega t+\phi\right)e^{-t/\tau_{\text{eff}}}+I_{\text{Land}}
\end{equation}
which, upon removal of the corresponding asymptotic Landauer current,
should give rise to the following complex components of the Fourier
transform: 
\begin{figure}[t]
\begin{centering}
\includegraphics[scale=0.245]{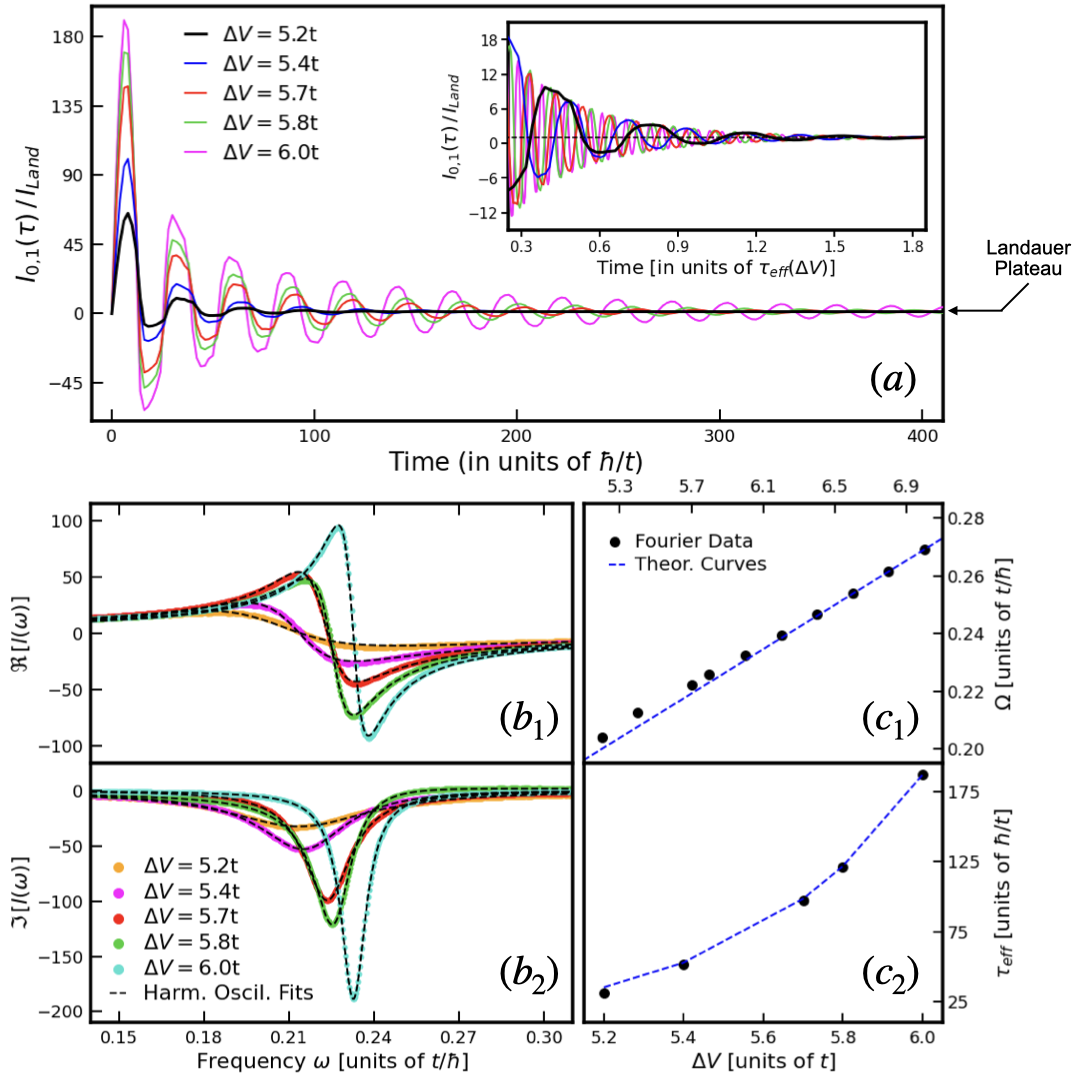}
\par\end{centering}
\centering{}\caption{\label{fig:FourierAnalysis}a) Average current evolution divided by
the respective Landauer value for different potential biases (in units
of $w$) for a sample of size 25 $\left(L=12\right)$. b) Same current
evolutions with time scaled by the inverse of HWHM coefficients and
amplitudes adjusted. The dashed brown line denotes Current/Landauer
= 1. b1) Real and b2) complex part of the Fourier transform of the
current for different potential biases (in units of $w$) for a system
of size 25 $\left(L=12\right)$. The dashed lines are the corresponding
fits of the functions in Eqs.\,(\ref{eq:asdfbre-1}) and (\ref{eq:asdfbre-2}).
c1) $\Omega$ values obtain from the fit for different $\Delta V$
(blue points) compared to the frequency of BOs for the corresponding
$\Delta V$ (dashed blue line). c2) $\tau_{\text{eff}}$ values obtain
from the fit for different $\Delta V$ (blue points) compared to the
corresponding values of $\tau_{0}\tau_{1}/\left(\tau_{0}+\tau_{1}\right)$
(dashed blue lines).}
\end{figure}

\begin{align}
\text{Re}\left[I_{0,1}\!\left(\omega\right)\right] & =A\tau_{\text{eff}}\frac{\cos\phi+\tau_{\text{eff}}\sin\phi\left(\omega\!-\!\Omega\right)}{1+\tau_{\text{eff}}^{2}\!\left(\omega-\Omega\right)^{2}}\label{eq:asdfbre-1}\\
\text{Im}\left[I_{0,1}\!\left(\omega\right)\right] & =A\tau_{\text{eff}}\frac{\sin\phi-\tau_{\text{eff}}\cos\phi\left(\omega-\Omega\right)}{1+\tau_{\text{eff}}^{2}\!\left(\omega\!-\!\Omega\right)^{2}}.\label{eq:asdfbre-2}
\end{align}

\noindent Having Eqs.\,(\ref{eq:asdfbre-1})-(\ref{eq:asdfbre-2})
as a template, we can now find the values of $\Omega$ and $\tau_{\text{eff}}$
by fitting the numerical data for $I_{0,1}\!(\omega)$ to these expressions
. The corresponding fits are presented in Fig\,\ref{fig:FourierAnalysis}(b)
and the values of $\Omega$ and $\tau_{\text{eff}}$ acquired for
various biases are shown in the panels of Fig\,\ref{fig:FourierAnalysis}(c).
From the presented results it is clear that:\textit{\,(i)} the oscillating
current is very well described as a single-frequency oscillation with
an exponentially decaying envelope, and \textit{(ii) }the values obtained
for the oscillation frequency and decay time perfectly agree with
the theoretical predictions of our strong bias model, i.e., $\Omega=2\pi/T_{\text{BO}}$
and $\tau_{\text{eff}}=\!\!\!\!\!=\!\tau_{0}\tau_{1}/\left(\tau_{0}\!+\!\tau_{1}\right)$.

\section{Conclusions and Outlook\label{sec:Conclusion}}

We have demonstrated how Bloch oscillations can coexist with the
formation of a steady-state current in biased mesoscopic devices.
Contrasting with Bloch oscillations in the Wannier-Stark model, these
oscillating current acquire a finite lifetime, due to the hybridization
to the device leads, and dub them transient Bloch oscillations. 

We performed a theoretical analysis of the occurrence of current Bloch
oscillations in a « one-dimensional mesoscopic system in a two-terminal
configuration. We focused on a nearest-neighbour tight-binding model,
with a partition-free initial condition and at half-filling, with
a constant electric field applied to its central region. We conclude
that Bloch oscillations can be observed provided $\ell_{\text{WS}}<La$,
where $\ell_{\text{WS}}=2w/\left(eEa\right)$ is the localization
length of WSS, with $w$ nearest-neighbour central region hopping
and $E$ the electric field that is applied to a region of length
$(2L+1)a$. In terms of applied bias voltage, this condition is equivalent
to $\Delta V>4w$. If the hopping in the leads is the same as the
hopping in the central region, the spectral bandwith of the leads
is $4w$, and the condition $\Delta V>4w$ implies that no steady-state
current can emerge, as previously found \citep{Popescu2017}. In this
regime, no current carrying scattering states can be constructed,
and bound states localized in the central region are observed. These
bound states are similar to the Wannier-Stark states of an infinite
chain subject to a constant electric field. For $\Delta V<4w$, a brief
build-up transient is followed by the emergence of a ballistic steady-state Landauer
current that flows through the device, but no Bloch oscillations are
observed. 
\begin{figure}[t]
\begin{centering}
\includegraphics[scale=0.42]{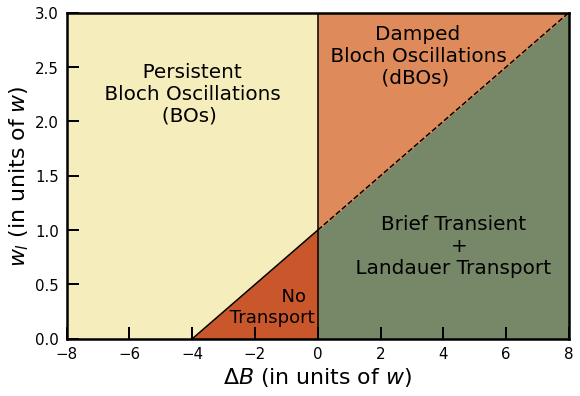}
\par\end{centering}
\begin{centering}
\caption{\label{phase}Diagram that summarizes the four different dynamical
phases of the strongly biased mesoscopic device, as a function of
the relative bandwidth of the leads, $w_{l}$, and the overlap between
the bands, $\Delta B\!=\!\delta\!\times\!\Delta V$. Four non-equilibrium
regimes are identified: \textit{(i)} persistent in-sample BOs, \textit{(ii)}\,tBOs
{[}decaying towards a Landauer steady-state current after, at least
one Bloch period{]}, \textit{(iii)}\,a Landauer transport steady-state,
and \textit{(iv)}\,a regime of blocked transport (with a chaotic
current inside the sample).}
\par\end{centering}
\vspace{-0.2cm}
\end{figure}

The aforementioned scenario in which Bloch oscillations and Landauer
steady-state transport are mutually exclusive quantum transport processes
gets drastically changed once the leads are permitted to have a wider
bandwidth than the sample. In particular, if the hoppings in the leads,
$w_{l}$, differs from the hoppings in the central region, $w$, is
is possible to observe the coexistence of Bloch oscillations with
the formation of a steady-state current provided $4w<\Delta V<4w_{l}$.
In this regime no true bound states localized in the central region
exist. Instead, we wave quasiparticle states, that resemble Wannier-Stark
states, but with a finite lifetime. As such, Bloch oscillations exist
as a transient phenomena, decaying in time until a constant and non-zero
steady-state current is formed. Crucially, these transient Bloch oscillations
display the same frequency of Bloch oscillations in a Wannier-Stark
ladder. Numerical simulations based on the unitary time-evolution
of the local electric current in a system coupled to finite leads,
in conjunction with a quasiparticle approximation scheme, were used
to demonstrate the validity of these claims. The \textit{phase diagram}
of Fig.\,\ref{phase}, which depicts the various dynamical phases
of the mesoscopic device as a function of the relative bandwidth of
the leads ($w_{l}$) and the overlap between the bands ($\Delta B=4w-\Delta V$)
summarizes the results. 

To summarize, we have found a new regime in which Bloch oscillations
can be observed as a transient phenomena in biased mesoscopic systems.
These Bloch oscillations could potentially be detected by the radiation
emitted by the oscillating current. 
\begin{acknowledgments}
Work supported by the Portuguese Foundation for Science and Technology
(FCT) within the Strategic Funding UIDB/04650/2020 and through projects
No.\,POCI-01-0145-FEDER-028887 (J.P.S.P., S.M.J and J.M.V.P.L.),
No.\,CEECIND/02936/2017 and No.~EXPL/FIS-MAC/0953/2021 (B.A.). J.P.S.P.
and S.M.J are funded by FCT grants No.\,PD/BD/142774/2018 and PD/BD/142798/2018,
respectively.
\end{acknowledgments}

\appendix

\section{Bloch Oscillations and the Wannier-Stark States\label{sec:Bloch-Oscillations-and}}

At the start of Sec.\,\ref{sec:Transport-Setup-and}, we revised
essential aspects of the exact solution for the tight-binding chain
subject to an uniform electric field. This turned out to be a crucial
theoretical cornerstone for our study because the shape of the WSSs
greatly aided in our comprehension of the various regimes of current
dynamics in a strongly biassed mesoscopic device. We did not, however,
fully examine all ramifications of this exact solution, particularly
how it relates to the presence of Bloch oscillations in this model.
For the sake of completeness, we provide more thorough discussion
in this appendix, referring to Hartmann \textsl{et al}.\,\citep{Hartmann2004}
for an in-depth approach.

Like before, we start from the Hamiltonian of the system in the presence
of a longitudinal electric field $E$, which reads,

\begin{equation}
\mathcal{H}_{{\scriptscriptstyle \text{WS}}}=\sum_{n=-\infty}^{+\infty}\left(-w\ket n\bra{n+1}+\ket{n+1}\bra n\!+aeEn\ket n\bra n\right),\label{eq:wavepacketham-1}
\end{equation}
$\ket n$ being local orbitals, $w$ the nearest-neighbor hopping,
and $a$ the lattice parameter. We have seen that the spectrum of
$\mathcal{H}_{{\scriptscriptstyle \text{WS}}}$ forms a so-called
\textit{Wannier-Stark ladder} with discrete energy levels, $\varepsilon_{m}\!=\!maeE$
(for $m\in\mathbb{Z}$), and that the corresponding eigenstates are
localized wavefunctions in real-space {[}quoted in Eq.\,(\ref{eq:WSS}){]}.
In place of repeating the real-space representation, we now highlight
that the WSSs can also be nicely represented in momentum space as
follows\,\citep{Hartmann2004}:

\begin{equation}
\ket{\Psi_{m}}=\sqrt{\frac{a}{2\pi}}\sum_{k}\exp\left[-iamk+\frac{2w}{iaeE}\sin ka\right]\ket{\phi_{k}},\label{eq:WannierStarkStatesMomentumSpace}
\end{equation}
where $\ket{\phi_{k}}$ are the lattice momentum eigenstates with
$-\pi\leq ka<\pi$. Using the eigenstates of the full Hamiltonian,
we can write down the exact time-evolution operator,

\begin{multline}
\mathcal{U}(t)=\int_{-\frac{\pi}{a}}^{\frac{\pi}{a}}dk\exp\left[\frac{2w}{iaeE}\sin\left(ka-\frac{eaEt}{\hbar}\right)\right]\\
\times\exp\left[-\frac{2w}{iaeE}\sin ka\right]\ket{\phi_{k-eEt/\hbar}}\bra{\phi_{k}},\label{eq:time_evvol_operator}
\end{multline}
where $t$ is the time parameter, and which can now be used to determine
the dynamics of any quantum state in which this system may start.
For example, if it starts from a thermal state, in the absence of
the electric field, as described by the reduced density matrix

\begin{equation}
\rho_{0}=\int_{-\frac{\pi}{a}}^{\frac{\pi}{a}}dkf\left(\varepsilon_{k}\right)\ket{\phi_{k}}\bra{\phi_{k}},
\end{equation}
where $f(\varepsilon)$ is the Fermi-Dirac distribution function,
and $\varepsilon_{k}=-2w\cos ka$ are the energy eigenvalues if $E=0$.
In such a case, the time-dependent expectation value of the total
electric current operator,

\begin{equation}
\begin{aligned}\mathcal{I} & =\frac{2eaw}{\hbar}\int_{-\frac{\pi}{a}}^{\frac{\pi}{a}}dk\sin ka\ket{\phi_{k}}\bra{\phi_{k}},\end{aligned}
\end{equation}
explicitly yields,

\begin{equation}
\begin{aligned}J(t) & =\text{Tr}\left[\rho_{0}\mathcal{U}(t)\mathcal{I}\mathcal{U}^{\dagger}(t)\right]\\
 & =-\frac{ea}{\hbar}\sin\left(\frac{eEat}{\hbar}\right)\int_{-\frac{\pi}{a}}^{\frac{\pi}{a}}dkf(\varepsilon_{k})\varepsilon_{k}.
\end{aligned}
\label{eq:Current}
\end{equation}
Surprisingly, Eq.\,(\ref{eq:Current}) demonstrates that, upon the
application of an uniform and static electric field, the electric
current oscillates in time with a period,

\begin{equation}
T_{{\scriptscriptstyle \textrm{BO}}}=\frac{2\pi\hbar}{aeE},
\end{equation}
that is inversely proportional to the applied electric field. 

\section{Appearance of a Beat Pattern in Bloch Oscillations\label{apx:BOs_modulation}}

\begin{figure}[b]
\begin{centering}
\includegraphics[width=1\columnwidth]{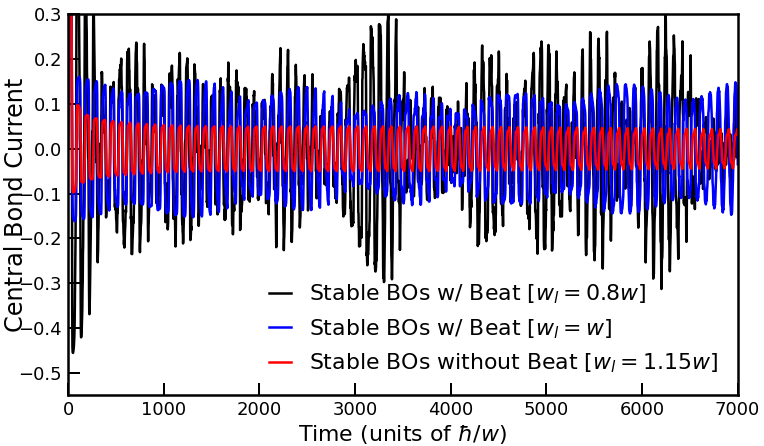}
\par\end{centering}
\caption{\label{fig:BeatOscilations}Plots of the local electric current measured
over time in the central bond of the mesoscopic sample composed of
69 sites ($L\!=\!34$) and a bias potential $\Delta V\!=\!5w$. We
showcase three examples of:\textit{ (i)} a mesoscopic device supporting
stable clipped BOs displaying a beat pattern with $w_{l}\!=\!0.8w$
\textit{(ii)} a mesoscopic device supporting stable clipped BOs displaying
a beat pattern with $w_{l}\!=\!w$ and \textit{(iii)} a mesoscopic
device supporting stable clipped BOs with no beat pattern with $w_{l}\!=\!1.15w$. }
\end{figure}

In figure (\ref{fig:DecayingOscilations}) we showcased stable BOs
and tBOs by changing the leads hoppings appropriately. For the sake
of brevity, we have omitted another effect from the main text, which
we will explain in this appendix instead. By setting $w_{l}=w$, the
originated BOs display a beat. This effect points to the introduction
of new time scales other than the Bloch period $T_{BO}$. Such time
scales appear due to the shift in the energy of the states centered
near the boundaries of the sample. This shifts makes it so that the
energetic difference between neihgbouring states is no longer equal
to a multiple of $eEa$. However, by increasing $w_{l}$, we can get
these states to couple to propagating ones in the leads, thereby allowing
them to escape the sample and making them not contribute to the central
current. Thus, setting $w_{l}$ to a sufficient high value but still
below $\Delta V/4$ shall eliminate the beat while maintaining the
BOs stable. In figure (\ref{fig:BeatOscilations}) we show the central
current evolution for three different values of $w_{l}$ below the
tBO threshold $\Delta V/4$: $0.8w$, $w$ and $1.15w$. We notice
that the beat is supressed with the increase of $w_{l}$, corroborating
our hypothesis.

\section{Study of Functions $C_{n}\left(\delta,\Delta V\right)$\label{sec:Study-of-Function}}

\begin{figure}[t]
\begin{centering}
\includegraphics[width=1\columnwidth]{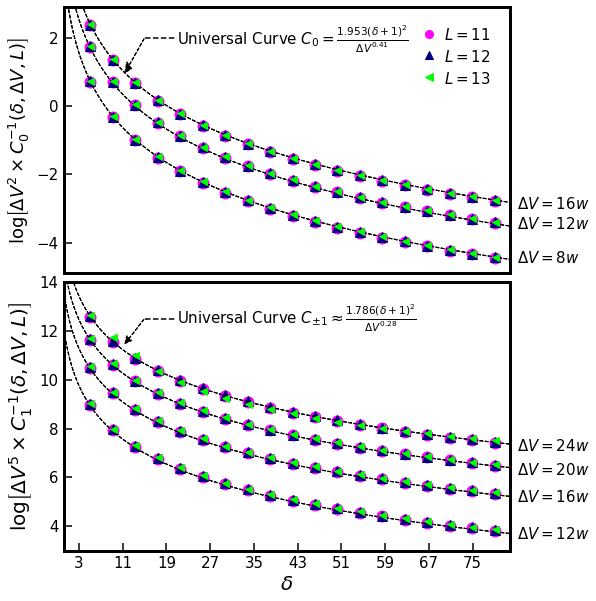}
\par\end{centering}
\centering{}\caption{\label{fig:cfunction}Plots of the inverse of the functions $C_{0}\left(\delta,\Delta V,L\right)$
and $C_{1}\left(\delta,\Delta V,L\right)$ (multiplied by $\Delta V^{2}$
and $\Delta V^{5}$ respectively for visualization purposes) as a
function of $\delta$ for different values of $\Delta V$ and $L$
(top and bottom panel respectively). The obtained points correspond
very well to the universal curves $C_{n}\left(\delta,\Delta V\right)=A_{n}\left(\delta+1\right)^{2}\Delta V^{\nu_{n}}$.}
\end{figure}

In subsection (\ref{subsec:Quasianalytic-estimation-of}), we have
said that the functions $C_{n}\left(\delta,\Delta V\right)$ have
a fixed dependence on $\delta$, a weak power law dependence on $\Delta V$
and do not depend on $L$. To back our claim, we plot these functions
for different values of $L$, $\delta$ and $\Delta V$ in figure
(\ref{fig:cfunction}). It is clear that these functions do not in
fact depend on $L$ and their behavior is congruent with universal
curves of the form $C_{n}\left(\delta,\Delta V\right)=A_{n}\left(\delta+1\right)^{2}\Delta V^{\nu_{n}}$.
The coefficients $A_{n}$ and $\nu_{n}$ are obtained from these plots.

\bibliographystyle{apsrev4-2}
\bibliography{refs}

\end{document}